\documentclass [12pt] {article}
\usepackage{latexsym}
\usepackage{latexsym}
\newtheorem{theorem}{Theorem}
\newtheorem{proposition}{Proposition}

\newtheorem{lemma}{Lemma}

\def\L{{\rm LZ}}

\title{Instability of Probability Laws with Respect to Small
Violations of Algorithmic Randomness}
\author{Vladimir V. V'yugin\thanks{
This paper is an extended version of the conference paper 
V'yugin~\cite{Vyu2011}.
}
\\
{\small Institute for Information Transmission Problems}\\
{\small Bol'shoi Karetnyi per. 19, Moscow GSP-4, 127994, Russia}\\
{\small e-mail vyugin@iitp.ru}
}

\date{}

\begin{document}
\maketitle

\begin{abstract}
We study a stability property of probability laws with respect to small
violations of algorithmic randomness. A sufficient condition of stability
of a probability law is presented in terms of Schnorr tests of algorithmic
randomness. Most probability laws, like the strong law of large numbers and
the law of iterated logarithm, are stable in this sense.

The phenomenon of instability takes place in ergodic theory. 
An instability of the Birkhoff's ergodic
theorem and related asymptotic laws with respect to small violations of
algorithmic randomness is proved in this paper. 
The Shannon--McMillan--Breiman theorem and
all universal compression schemes are also instable.
\end{abstract}

\section{Introduction}

In this paper we study a stability property of probability laws
with respect to small violations of randomness. By a
probability law we mean a property $\Phi(\omega)$ of infinite
sequences $\omega$ which holds almost surely. We define a
notion of stability of a probability law in terms of algorithmic theory of
randomness. Within the framework of this theory the probability
laws are formulated in ``a pointwise'' form.
%This means that such a law holds not only
%almost surely but also for each Martin-L\"of random sequence.
It is well known that main laws of probability theory are valid
not only almost surely but for each individual Martin-L\"of random sequence.

Some standard notions on algorithmic randomness are given in Section~\ref{random-1}.
We use the definition of a random sequence in the complexity terms.
An infinite binary sequence $\omega_1\omega_2\dots$ is
Martin-L\"of random with respect to uniform (or $1/2$-Bernoulli) measure
if and only if $Km(\omega^n)\ge n-O(1)$ as $n\to\infty$,
where $Km(\omega^n)$ is a monotone Kolmogorov complexity of
a binary string $\omega^n=\omega_1\dots\omega_n$ and the constant $O(1)$
depends on $\omega$ but not on $n$.
\footnote
{The same property holds also if we replace monotonic complexity $Km(\omega^n)$
on the prefix complexity $KP(\omega^n)$. A difference is that an inequality
$Km(\omega^n)\le n+O(1)$ holds for monotonic complexity 
but this is not true for prefix complexity.}

A probability law $\Phi(\omega)$ is called stable if an unbounded
computable function
$\rho(n)$ exists such that $\Phi(\omega)$ is true for any infinite
sequence $\omega$ such that $Km(\omega^n)\ge n-\rho(n)-O(1)$ as $n\to\infty$.

A stability property under small violations of algorithmic randomness
of the main limit probability laws was discovered by Schnorr~\cite{Sch71} and
Vovk~\cite{Vov87}. They shown that the law of large numbers for
uniform Bernoulli measure holds for a binary sequence $\omega_1\omega_2\dots$
if $Km(\omega^n)\ge n-\sigma(n)-O(1)$, where $\sigma(n)$ is an arbitrary computable
function such that $\sigma(n)=o(n)$ as $n\to\infty$, and
the law of iterated logarithm holds if $Km(\omega^n)\ge n-\sigma(n)-O(1)$,
where $\sigma(n)$ is an arbitrary computable function such that
$\sigma(n)=o(\log\log n)$ as $n\to\infty$.
\footnote
{
All logarithms are on the base 2.
}

We present in Theorem~\ref{suff-1} a sufficient condition of stability
in terms of Schnorr tests of randomness and
prove that most probability laws, like the
strong law of large numbers and the law of iterated logarithm, are stable
under small violations of algorithmic randomness.

In Section~\ref{instab-1} we show that the situation is different
in ergodic theory. We note that there is some analogy
with the lack of universal convergence rate estimates in ergodic theory.
A lack of universal convergence bounds is typical for
asymptotic results of ergodic theory like Birkhoff's ergodic
theorem -- Krengel~\cite{Kre85}, Shannon--McMillan--Breiman
theorem and universal compressing schemes --
Ryabko~\cite{Ryb84}.

In this paper we present new impossibility results based on
analysis of ergodic theory in algorithmic randomness framework
-- instability under small violations of algorithmic
randomness.

Some notions of ergodic theory are given in Section~\ref{cat-sta}.
Let $T$ be a measure preserving transformation and $f$ be a real-valued
function defined on the set of binary sequences.

We suppose that the transformation $T$ and the function $f$ are computable
(see definitions in Section~\ref{random-1}).

Using Bishop's~\cite{Bis67} analysis, V'yugin~\cite{Vyu98}
presented an algorithmic version of the ergodic theorem --
for any infinite binary sequence $\omega$ the following implication is valid~:
\begin{eqnarray}
Km(\omega^n)\ge n-O(1)\Longrightarrow\lim\limits_{n\to\infty}\frac{1}{n}
\sum\limits_{i=0}^{n-1}f(T^i\omega)=\hat f(\omega)
\label{e-1}
\end{eqnarray}
for some $\hat f(\omega)$ ($=E(f)$ for ergodic $T$).

Later this result was extended for non-computable $f$ and generalized
for more general metric spaces.
For further development see Nandakumar~\cite{Nan2008},
Galatolo et al.~\cite{GHR2010}, and P. Gacs et al.~\cite{GHR2011}.

We show in Section~\ref{instab-1} an instability of this asymptotic
result, more correctly, that a constant $O(1)$ in (\ref{e-1}) cannot be
replaced on an arbitrary slow increasing computable variable $\sigma(n)$.

\section{Stability of probability laws}\label{stab-1}

Let $\Omega=\{0,1\}^\infty$ be the set of all infinite
binary sequences. In what follows $l(\alpha)$ denotes the length
of a finite sequence $\alpha$. The uniform Bernoulli measure is defined
$L(\Gamma_\alpha)=2^{-l(\alpha)}$,
where $\Gamma_\alpha=\{\omega\in\Omega:\alpha\subseteq\omega\}$
for any finite binary string $\alpha$,.

In Section~\ref{random-1} the notions of monotonic complexity
$Km(x)$ and of Martin-L\"of and Schnorr random sequences are defined.

Let $\Phi(\omega)$ be an asymptotic probability law, i.e., a property
of infinite binary sequences which holds almost surely.

Kolmogorov's algorithmic approach to probability
theory offers a new paradigm for logic of probability. The requirement
``$\Phi(\omega)$ holds almost surely'' is replaced on a more strong
requirement:

``$\omega$ is Martin-L\"of random $\Longrightarrow$ $\Phi(\omega)$''.

For any infinite binary string $\omega$,
$Km(\omega^n)\le n+O(1)$ for all $n$. The converse inequality holds
for any Martin-L\"of random sequence $\omega$. Namely,
an infinite sequence $\omega=\omega_1\omega_2\dots$ is Martin-L\"of random
if and only if $Km(\omega^n)=n+O(1)$ as $n\to\infty$
(see Section~\ref{random-1}).

Therefore, we can formulate an equivalent form of a probabilistic law:

``$Km(\omega^n)\ge n-O(1)$ as $n\to\infty$ $\Longrightarrow$ $\Phi(\omega)$''

In this paper we present a more deep analysis. We call a law
$\Phi(\omega)$ stable if there exists a unbounded nondecreasing
computable function $\alpha(n)$ such that the following property holds:

``$Km(\omega^n)\ge n-\alpha(n)-O(1)$ as $n\to\infty$ $\Longrightarrow$ $\Phi(\omega)$''

The function $\alpha(n)$ is called degree of stability of the law
$\Phi(\omega)$.

We present in this paper a sufficient condition for stability
of a probability law and prove that most probability laws are
stable with different degrees of stability. We formulate this condition
in terms of Schnorr~\cite{Sch71} definition of algorithmic random sequence.
See Section~\ref{random-1} for details of this definition.
The choice of Schnorr's definition is justified by an observation that
for any probability law $\Phi(\omega)$ from a vast majority of such laws 
(like the strong the law of large numbers or the law of iterated logarithm)
a computable sequence ${\cal T}_n$, $n=1,2,\dots$,
of Schnorr tests of randomness exists such that for any infinite sequence
$\omega$, $\Phi(\omega)$ is true if and only if $\omega$ passes 
the test ${\cal T}_k$ for each $k$.

For technical reasons we consider total Solovay tests of randomness
which leads to the same definition of randomness as Schnorr tests of randomness.
A computable sequence ${\cal T}=\{x_n:n=1,2,\dots\}$ of finite strings is called
total Solovay test of randomness if the series
$\sum\limits_{n=1}^\infty 2^{-l(x_n)}$ converges with
computable rate of convergence. 

An infinite sequence $\omega$ passes a test $\cal T$ if 
$x_n\not\subseteq\omega$ for almost all $n$
(see a more detailed consideration in Section~\ref{random-1}).

In the following theorem some sufficient condition of stability of
a probability law is given in terms of total Solovay tests randomness.

By computable sequence of total Solovay tests of randomness we mean
a computable double indexed sequence of finite binary
strings ${\cal T}_k=\{x_{k,n}:n=1,2,\dots\}$, $k=1,2,\dots$, such that the series
$\sum\limits_{n=1}^\infty 2^{-l(x_{k,n})}$
converges with a uniformly by $k$ computable rate of convergence.
This means that
%in (\ref{coverg-1})
there exists a computable function
$m(\delta,k)$ such that
$\sum\limits_{i=m(\delta,k)}^\infty 2^{-l(x_{k,i})}\le\delta$ for each $k$
and rational $\delta$.

\begin{theorem}\label{suff-1}
For any computable sequence of total Solovay tests of randomness
${\cal T}_k$, $k=1,2,\dots$, a computable unbounded function $\rho(n)$ exists
such that for any infinite sequence $\omega$ if
$Km(\omega^n)\ge n-\rho(n)-O(1)$ as $n\to\infty$ then the sequence $\omega$
passes the test ${\cal T}_k$.
\end{theorem}
{\it Proof}. Let ${\cal T}_k=\{x_{k,n}:n=1,2,\dots\}$ for any $k=1,2,\dots$.
Since
$$
\sum\limits_{m=1}^\infty 2^{-l(x_{k,m})}<\infty
$$
with a uniform computable rate of convergence, an unbounded nondecreasing
computable function $\nu(n)$ exists such that
$$
\sum\limits_{m=1}^\infty 2^{-l(x_{k,m})+\nu(l(x_{k,m}))}<\infty
$$
for any $k$.
By generalized Kraft inequality (see Li and Vitanyi~\cite{LiV97}), 
for any $k$, we can define the corresponding prefix-free code uniformly
computable by $k$ such that
$$
Km(x_{k,m})\le l(x_{k,m})-\nu(l(x_{k,m}))+O(1)
$$
Assume an $k$ exists such that $x_{k,m}\subseteq\omega$ for
infinitely many $m$. For any such $m$, $\omega^n=x_{k,m}$, where $n=l(x_{k,m})$.

Let $\rho(n)$ be a unbounded nondecreasing computable function such that
$\rho(n)=o(\nu(n))$ as $n\to\infty$. Let also, $\omega$ be an infinite
binary sequence such that $Km(\omega^n)\ge n-\rho(n)-O(1)$ for all $n$.
For some $k$ and $n=l(x_{k,m})$,
\begin{eqnarray*}
\rho(n)\ge n-Km(\omega^n)\ge
%\nonumber
%\\
n-l(x_{k,m})+\nu(l(x_{k,m}))-O(1)
%\nonumber
%\\
\ge 
%n-n+\nu(n)-O(1)=
\nu(n)-O(1)
\end{eqnarray*}
for infinitely many $n$. On the other hand,
$\rho(n)=o(\nu(n))$ as $n\to\infty$.
This contradiction proves the theorem.
$\triangle$

Using Theorem~\ref{suff-1} we can prove the stability property
for main probability laws.

At first, we show that the strong law of large numbers
corresponds to an infinite sequence of total Solovay randomness tests.

Hoeffding~\cite{Hoe63} inequality for uniform probability distribution $L$
says that for any $\epsilon>0$
\begin{eqnarray}
L\left\{\omega\in\Omega: \left|\frac{1}{n}\sum\limits_{i=1}^n
\omega_i-\frac{1}{2}\right|\ge\epsilon\right\}\le 2e^{-2n\epsilon^2}
\label{strong-1g}
\end{eqnarray}
for all $n$. 

Let $\epsilon_k$ be a computable sequence of positive rational numbers 
such that $\epsilon_k\to 0$ as $k\to\infty$. Define a sequence of sets
\begin{eqnarray}
U_{k,n}=\left\{x: l(x)=n,\left|\frac{1}{n}\sum\limits_{i=1}^n
x_i-\frac{1}{2}\right|\ge\epsilon_k\right\}.
\label{strong-1}
\end{eqnarray}
By (\ref{strong-1g})
\begin{eqnarray}
L(U_{k,n})=\sum\limits_{x\in U_{k,n}}2^{-l(x)}\le 2e^{-2n\epsilon^2_k}
\label{hoeff-1}
\end{eqnarray}
for all $n$ and $k$.

Let $\cup_n U_{k,n}=\{x_{k,m}:m=1,2,\dots\}$.
For any $k$, the sequence 
$$
\{x_{k,m}:m=1,2,\dots\}
$$ 
is a total Solovay test of randomness, since
\begin{eqnarray*}
\sum\limits_{m=1}^\infty 2^{-l(x_{k,m})}=
\sum\limits_{n=1}^\infty L(U_{k,n})\le
\sum\limits_{n=1}^\infty 2e^{-2n\epsilon^2_k}<\infty
\end{eqnarray*}
with a computable rate of convergence.

By definition (\ref{strong-1}) the strong law of large numbers
$$
\lim\limits_{n\to\infty}\frac{1}{n}\sum\limits_{i=1}^n\omega_i=\frac{1}{2}
$$
holds for an infinite sequence $\omega=\omega_1\omega_2\dots$
if and only if for each $k$ it passes the test $\{x_{k,m}:m=1,2,\dots\}$.

Then by Theorem~\ref{suff-1} an unbounded nondecreasing
computable function $\rho(n)$ exists
such that if $Km(\omega^n)\ge n-\rho(n)-O(1)$ as $n\to\infty$ then for any $k$
the sequence $\omega$ passes the test $\{x_{k,m}:m=1,2,\dots\}$.
This means that for any $k$, $\omega^n\not\in U_{k,n}$ for almost all $n$,
i.e., the strong law of large numbers (\ref{strong-1}) holds for this $\omega$.

We can find the specific form of this function $\rho(n)$.
Consider any unbounded nondecreasing computable function
$\rho(n)$ such that $\rho(n)=o(n)$ as $n\to\infty$. Let also,
$\epsilon_k$ be a computable sequence of positive rational numbers 
such that $\epsilon_k\to 0$ as $k\to\infty$. It is easy
to see that an unbounded nondecreasing computable function
$\alpha(n)$ exists such that $\alpha(n)=o(n)$ and
$\rho(n)=o(\alpha(n))$ as $n\to\infty$.

We write $\L(U_{k,n})\le 2^{-\alpha(n)}c_{k,n}$, where
$c_{k,n}=2e^{-2n\epsilon^2_k+\alpha(n)}$. For any $k$,
\begin{eqnarray*}
\sum\limits_{n=1}^\infty 2^{-l(x_{k,n})+\alpha(l(x_{k,n}))}=
\sum\limits_{n=1}^\infty 2^{\alpha(n)}L(U_{k,n})\le
\sum\limits_{n=1}^\infty c_{k,n}<\infty.
\end{eqnarray*}
since $\alpha(n)=o(n)$ as $n\to\infty$.
By the proof of Theorem~\ref{suff-1} if $Km(\omega^n)\ge n-\rho(n)$ then
for each $k$,
$\omega^n\not\in U_{n,k}$ for almost all $n$, i.e., the strong law of
large numbers holds for this $\omega$.

An analogous construction can be developed for the law of iterated logarithm:
\begin{eqnarray}
\limsup\limits_{n\to\infty}\frac{\left|\sum\limits_{i=1}^n\omega_i-\frac{n}{2}\right|}
{\sqrt{\frac{1}{2}n\ln\ln n}}=1.
\label{iterr-1}
\end{eqnarray}
We prove here only the inequality $\le$ in (\ref{iterr-1}).
%analyze a construction from~Shiryaev~\cite{Shi80}, Chapter IV.
This inequality violates if and only if a rational number
$\delta>1$ exists such that
$$
S_n-\frac{n}{2}>\delta\sqrt{\frac{1}{2}n\ln\ln n}
$$
for infinitely many $n$, where $S_n=\sum\limits_{i=1}^n\omega_i$.

For any rational number $\delta$ such that $\delta>1$ and for
$m_{n}=\lceil\delta^{n}\rceil$, let
\footnote{For any real number $r$, $\lceil r\rceil$ denotes the least
positive integer number $m$ such that $m\ge r$.}

\begin{eqnarray}
U_{\delta,n}=
\nonumber
\\
=\{\omega\in\Omega: \exists k(m_{n}\le k\le m_{n+1}\&
%\nonumber
%\\
S_k-k/2>\delta\sqrt{(1/2)m_n\ln\ln m_n}\}.
\label{iterr-2}
\end{eqnarray}
Using the inequality
$$
L\{\max\limits_{1\le k\le m} S_k>a\}\le 2 L\{S_m>a\},
$$
we obtain 
\begin{eqnarray}
L(U_{\delta,n})\le 
\nonumber
\\
\le L(\{\omega\in\Omega: \exists k(k\le m_{n+1}\&S_k-k/2>
\delta\sqrt{(1/2)m_{n}\ln\ln m_{n}})\})\le
\nonumber
\\
\le 2 L(\{\omega\in\Omega: S_{m_{n+1}}-m_{n+1}/2>
\delta\sqrt{(1/2)m_{n}\ln\ln m_{n}}\})\le
\nonumber
\\
\le ce^{-\delta\ln\ln m_{n}}\approx\frac{1}{n^\delta},
\label{iterr-3}
\end{eqnarray}
where $c>0$. We have used in (\ref{iterr-3}) the 
Hoeffding inequality (\ref{hoeff-1}).

We can effectively construct a prefix-free set $\tilde U_{\delta,n}$
of finite sequences such that for each $\omega\in U_{\delta,n}$
an $m$ exists such that $\omega^m\in\tilde U_{\delta,n}$.

A sequence
$\cup_n\tilde U_{\delta,n}=\{x_{\delta,k}:k=1,2,\dots\}$
is a total Solovay test of randomness, since the series
$$
\sum\limits_n 2^{-l(x_{\delta,n})}=\sum\limits_n L(U_{\delta,n})\le
\sum\limits_n \frac{1}{n^\delta}
$$
converges (with a computable rate of convergence) for any $\delta>1$.

By definition the law of iterated logarithm (\ref{iterr-1}) holds
for $\omega=\omega_1\omega_2\dots$
if and only if it passes the test $\{x_{\delta,k}:k=1,2,\dots\}$
for each $\delta>1$.

By Theorem~\ref{suff-1} an unbounded nondecreasing
computable function $\rho(m)$ exists such that the inequality $\le$ in 
(\ref{iterr-1}) holds
for any $\omega$ satisfying $Km(\omega^m)\ge m-\rho(m)-O(1)$ as $m\to\infty$.

We can also find a specific form of $\rho(m)$ for the law of iterated logarithm.
Let $\rho(m)$ be a unbounded nondecreasing computable
function $\rho(m)$ such that $\rho(m)=o(\log\log m)$ as $m\to\infty$.
An unbounded nondecreasing computable
function $\alpha(m)$ exists such that $\alpha(m)=o(\log\log m)$
and $\rho(m)=o(\alpha(m))$ as $m\to\infty$.

Since $\alpha(m_n)=o(\ln\ln m_n)=o(\ln n)$, the series
$$
\sum\limits_n e^{-\delta\ln\ln m_{n+1}+\alpha(m_n)}\approx
\sum\limits_n\frac{o(\ln n)}{n^\delta}
$$
converges for any $\delta>1$.
By the proof of Theorem~\ref{suff-1},
any sequence $\omega$ passes the test $\{x_{\delta,n}:n=1,2,\dots\}$
if $Km(\omega^m)\ge m-\rho(m)-O(1)$ as $m\to\infty$.

\section{Non-stable probability laws}\label{instab-1}

The phenomenon of instability takes place in ergodic theory. 
Some basic notions of ergodic theory are given in Section~\ref{cat-sta}.
In this section we present an instability property of the ergodic theorem.

In Theorems~\ref{th-1} and~\ref{th-1b} the uniform measure
$L$ and measure preserving transformations of $\Omega$ are considered.
\begin{theorem}\label{th-1}
Let $\sigma(n)$ be a nondecreasing unbounded computable function. Then there
exist a computable ergodic measure preserving transformation $T$
and a sequence $\omega\in\Omega$
such that the inequality $Km(\omega^n)\ge n-\sigma(n)$ holds
for all $n$ and the limit
\begin{eqnarray}
\lim\limits_{n\to\infty}\frac{1}{n}
\sum\limits_{i=0}^{n-1}f(T^i\omega)
\label{e-1a}
\end{eqnarray}
does not exist for some
computable indicator function $f$ (with the range $\{0,1\}$).
\end{theorem}
%``a measure free'' (comparing with~\cite{Vyu2003})
The construction of the transformation $T$ is given in Section~\ref{sec-1};
the proof of Theorem~\ref{th-1} is given in Section~\ref{sec-2}.

In the following theorem an uniform with respect to $\sigma(n)$
result is presented. But in this case, we loss the ergodic property of
the transformation $T$.
\begin{theorem}\label{th-1b}
A measure preserving transformation $T$ can be constructed such that
for any nondecreasing unbounded computable function $\sigma(n)$
a sequence $\omega\in\Omega$ exists such that $Km(\omega^n)\ge n-\sigma(n)$
holds for all $n$ and the limit (\ref{e-1}) does not exist for some
computable indicator function $f$.
\end{theorem}
On the proof of this theorem see Section~\ref{sec-2}.

Does an ergodic transformation $T$ exists satisfying Theorem~\ref{th-1b}
is an open question.

%Let $T$ be the left shift on $\Omega$, and $P$ be the computable
%stationary measure (i.e. the left shift $T$ is invariant w.r.t $P$)
%defined by (\ref{mes-1a}) using the partition $(\pi_0,\pi_1)$.

Note that an infinite sequence $\omega$ is Martion-L\"of random with respect to
a computable measure $P$ if and only if
$$
Km(\omega^n)=-\log P(\omega^n)+O(1)
$$
as $n\to\infty$.

Recent result of Hochman~\cite{Hoch2009} implies
an algorithmic version of the Shannon--McMillan--Breiman theorem
for Martin-L\"of random sequences:
for any computable stationary ergodic measure $P$ with entropy $H$,
$Km(\omega^n)\ge -\log P(\omega^n)-O(1)$ as $n\to\infty$ implies
\begin{eqnarray}
\lim\limits_{n\to\infty}\frac{Km(\omega^n)}{n}=
\lim\limits_{n\to\infty}\frac{-\log P(\omega^n)}{n}=H
\label{kol-comp-1}
\end{eqnarray}

The construction given in Section~\ref{sec-1} shows also an instability property
of the relation (\ref{kol-comp-1}) (this was first shown in~\cite{Vyu2003}).
\begin{theorem} \label{theorem-1}
Let $\sigma(n)$ as in Theorem~\ref{th-1}, $\epsilon$ be a sufficiently small
positive real number. A computable stationary ergodic measure $P$
with entropy $0<H\le\epsilon$ and an infinite binary sequence $\omega$
exist such that
\begin{eqnarray}
Km(\omega^n)\ge-\log P(\omega^n)-\sigma(n)
\label{def-lim-1}
\end{eqnarray}
for all n, and
\begin{eqnarray}
\limsup\limits_{n\to\infty}\frac{Km(\omega^n)}{n}\ge\frac{1}{4},
\label{limsup-1}
\\
\liminf\limits_{n\to\infty}\frac{Km(\omega^n)}{n}\le\epsilon.
\label{liminf-1}
\end{eqnarray}
\end{theorem}
The proof of this theorem is given in Section~\ref{sec-3}.

By a prefix-free code we mean a computable sequence of one-to-one functions
$\{\phi_n\}$ from $\{0,1\}^n$ to a prefix-free set of finite sequences.
In this case a decoding method $\hat\phi_n$ also exists such that
$\hat\phi_n(\phi_n(\alpha))=\alpha$ for each $\alpha$ of length $n$.

A code $\{\phi_n\}$ is called {\it universal coding scheme} with respect to a class
of stationary ergodic sources if for any computable stationary ergodic
measure $P$ (with entropy $H$)
\begin{eqnarray}\label{asym-zv}
\lim_{n\to\infty}\frac{l(\phi_n(\omega^{n}))}{n}=H\mbox{ a.s., }
\end{eqnarray}
where $l(x)$ is length of a finite sequence $x$.

Lempel--Ziv coding scheme is an example of such universal coding scheme.

We have also an instability property for any universal coding schemes.

\begin{theorem}\label{theorem-2}
Let $\sigma(n)$ and $\epsilon$ be as in Theorem~\ref{th-1}.
A computable stationary ergodic measure $P$ with entropy $0<H\le\epsilon$
exists such that for each universal code $\{\phi_n\}$ an infinite binary
sequence $\omega$ exists such that
\begin{eqnarray*}
Km(\omega^n)\ge-\log P(\omega^n)-\sigma(n)
\end{eqnarray*}
for all $n$, and
\begin{eqnarray}
%\limsup\limits_{n\to\infty}\rho_{\phi_n}(\omega^{n})\ge\frac{1}{4},
\limsup\limits_{n\to\infty}\frac{l(\phi_n(\omega^{n}))}{n}\ge\frac{1}{4},
\label{limsup-1mb}
\\
%\liminf\limits_{n\to\infty}\rho_{\phi_n}(\omega^{n})\le\epsilon
\liminf\limits_{n\to\infty}\frac{l(\phi_n(\omega^{n}))}{n}\le\epsilon.
\label{liminf-1mb}
\end{eqnarray}
\end{theorem}
The proof of this theorem is given in Section~\ref{sec-3}.

%\section{Proof of Theorem~\ref{th-1b}}~\label{sec-1}

\section{Construction}~\label{sec-1}

Basic notions of algorithmic randomness theory and ergodic
theory are given in Sections~\ref{random-1} and~\ref{cat-sta}
correspondingly.

We use a cutting and stacking method from 
Shields~\cite{Shi91}~and~\cite{Shi93}
to define an ergodic transformation
$T$ of the unit interval $[0,1)$. See Section~\ref{cat-sta} for
all needed definitions and details of this method.

Let $r>0$ be a sufficiently small rational number. Define a partition
\begin{eqnarray}
\pi_0=[0,0.5)\cup (0.5+r,1),\mbox{ }
\pi_1=[0.5,0.5+r]
\label{partition-1}
\end{eqnarray}
of the interval $[0,1)$ (the number $r$ will be specified later).
We consider the uniform measure $\lambda$ on $[0,1)$.

Let $\sigma(n)$ be a computable unbounded nondecreasing function.
A computable sequence of positive integer numbers
exists such that $0<h_{-2}<h_{-1}<h_0<h_1<\dots$ and
\begin{equation}\label{speed-1}
%\sigma(h_{i-1})-\sigma(h_{i-2})>i-\log r+8
\sigma(h_{i-1})-\sigma(h_{i-2})>i-\log r+11
\end{equation}
for all $i=0,1,\dots$.

The gadgets $\Delta_{s}$, $\Pi_{s}$, where $s=0,1,\dots$,
will be defined by mathematical induction on steps.
The gadget $\Delta_{0}$ is defined by cutting of the interval
$[0.5-r,0.5+r]$ on $2h_0$ equal parts and by stacking them.
Let $\Pi_0$ be a gadget defined by cutting of the intervals
$[0,0.5-r)$ and $(0.5+r,1]$ in $2h_0$ equal subintervals
and stacking them. The purpose of this definition is to construct initial
gadgets of height $2h_0$ with supports satisfying
$\lambda(\hat\Delta_0)=2r$ and $\lambda(\hat\Pi_0)=1-2r$.

The sequence of gadgets $\{\Delta_{s}\}$, $s=0,1,\dots$, will
define an approximation of the uniform Bernoulli measure
concentrated on the names ot their trajectories. The sequence
of gadgets $\{\Pi_s\}$, $s=0,1,\dots$, will define a measure
with sufficiently small entropy.
The gadget $\Pi_{s-1}$ will be extended at each step of the
construction by a half part of the gadget $\Delta_{s-1}$. After that, the
independent cutting and stacking process will be applied to this extended
gadget. This process eventually defines infinite trajectories starting from
points of the interval $[0,1)$. The sequence of gadgets
$\{\Pi_s\}$, $s=0,1,\dots$, will be complete and will define
a transformation $T$ of the interval $[0,1)$
and a measure $P$ on these trajectories (defined by (\ref{mes-1a}).
Lemmas~\ref{well-def} and \ref{M-fold} from Section~\ref{cat-sta} ensure
the transformation $T$ and measure $P$ to be ergodic.

{\it Construction}.
Let at step $s-1$ ($s>0$) gadgets $\Delta_{s-1}$ and $\Pi_{s-1}$
were defined. Cut of the gadget $\Delta_{s-1}$ into two copies
$\Delta'$ and $\Delta''$ of equal width (i.e. we cut of each column into two
subcolumns of equal width) and join $\Pi_{s-1}\cup\Delta''$
in one gadget. Find a number $R_s$ and do $R_s$-fold independent cutting
and stacking of the gadget $\Pi_{s-1}\cup\Delta''$ and also of the
gadget $\Delta'$ to obtain new gadgets $\Pi_s$ and $\Delta_{s}$ of height
$\ge 2h_s$ such that the gadget $\Pi_{s-1}\cup\Delta^{''}$ is
$(1-1/s)$--well--distributed in the gadget $\Pi_s$. The needed
number $R_s$ exists by Lemma~\ref{M-fold} (Section~\ref{cat-sta}).

{\it Properties of the construction}. Define a transformation $T=T\{\Pi_s\}$
on the interval $[0,1)$.

Since the sequence of the gadgets $\{\Pi_s\}$ is complete
(i.e. $\lambda({\hat\Pi}_s)\to 1$ and $w(\Pi_s)\to 0$
as $s\to\infty$), $T$ is defined for $\lambda$-almost all points of the
unit interval.

The transformation $T$ is ergodic by Lemma~\ref{well-def}
(Section~\ref{app}), where $\Upsilon_s=\Pi_{s}$, since the sequence of gadgets
$\Pi_s$ is complete. Besides, the gadget $\Pi_{s-1}\cup\Delta''$,
and the gadget $\Pi_{s-1}$ are $(1-1/s)$-well--distributed
in $\Pi_s$ for any $s$. By construction
$\lambda(\hat\Delta_i)=2^{-i+1}r$ and $\lambda(\hat\Pi_i)=1-2^{-i+1}r$
for all $i=0,1,\dots$.

\section{Proof of Theorems~\ref{theorem-1}~and~\ref{theorem-2}}~\label{sec-3}

Recall some notions of symbolic dynamics. Let $\pi=(\pi_1,\pi_2)$
be a partition of the unit interval: $\pi_1$ and $\pi_2$ are disjoint and
$\pi_1\cup\pi_2=[0,1)$. 
A transformation $T$ of the interval $[0,1)$ defines a measure $P$ on
the set of all finite and infinite binary sequences:
\begin{equation} \label{mes-1a}
P(\Gamma_x)=\lambda\{\alpha\in [0,1) :
T^i \alpha\in\pi_{x_i},\mbox i=1,2,\dots,n\},
\end{equation}
where $\lambda$ is the uniform measure on $[0,1)$ and
$x=x_1\dots x_n$ is a binary sequence. We use notation
$P(x)=P(\Gamma_x)$.

The measure $P$ can be extended on all Borel subsets of $\Omega$
by the Kolmogorov's extension theorem.
The measure $P$ defined by (\ref{mes-1a}) is stationary and ergodic
with respect to the left shift if and only if $T$ is a measure
preserving ergodic transformation.

In the construction of Section~\ref{sec-1} we have defined
a specific partition $(\pi_0,\pi_1)$ of $[0,1)$ and
the the measure preserving ergodic transformation $T$ of $[0,1)$
using a sequence of gadgets
$\Delta_{s}$, $\Pi_{s}$, where $s=0,1,\dots$.
The union of these two gadgets is denoted $\Phi_s$.

At a step $s$, we consider an approximation $T_s=T\{\Phi_s\}$ of
the transformation $T$ and the corresponding approximation $P^s$
of the measure $P$ defined by (\ref{mes-1a}).
The transformation $T_s$ determines
finite trajectories starting in the points of internal intervals
of this gadget and finishing in the top intervals. Any such trajectory
has a name which is a word in the alphabet $\{0,1\}$.
By definition for any word $a$ (for any set of words $D$)
the number $P^s(a)$ ($P^s(D)$ accordingly) is equal to the sum
of lengths of all intervals of the gadget $\Pi_{s}$ from which
trajectories with names extending $a$ (extending a word from $D$) start.

We use the construction of Section~\ref{sec-1}
to suggest conditions under which there exists
a point in the interval $[0,1)$ having an infinite trajectory
with a name $\alpha$ satisfying~(\ref{def-lim-1}), (\ref{limsup-1})
and (\ref{liminf-1}). To implement~(\ref{limsup-1}), we periodically
extend initial fragments of $\alpha$ by names of trajectories
of gadgets $\Delta_{s-1}$ (for a suitable $s$). By the incompressibility
property most of them have Kolmogorov complexity close to its maximal value.

We use condition~(\ref{speed-1}) and Proposition~\ref{deff-1}
from Section~\ref{increase} to bound the deficiency of randomness
of each initial fragment $\alpha^n$ of length $n$ by the value $\sigma(n)$.

To implement condition~(\ref{liminf-1}), we extend the initial fragments
of $\alpha$ in the long runs of the construction only in account of names
of the trajectories of gadgets $\{\Pi_s\}$, $s=0,1,\dots$.
For any $s$, a portion $\le r$ of the support of such a gadget belongs
to the element $\pi_1$ of the partition. By the ergodic theorem the most
part of (sufficiently long) trajectories of this gadget visit $\pi_1$
according to this frequency. The names of these trajectories
have frequency of ones bounded by the number $2r$ from above. This
ensures the bound (\ref{liminf-1}) for a sufficiently small $r$.

This construction is algorithmic effective, so the measure $P$ is computable.

Let $\epsilon>0$ be given. Now we prove that entropy $H$ of the measure $P$
is less than $\epsilon$ if $r$ is sufficiently small.
Since $\lambda(\pi_1)=r$ and the transformation $T$ preserves the measure
$\lambda$, by the ergodic theorem almost all points of the interval
$[0,1)$ generate trajectories with the frequency $r$ of visiting the
element $\pi_1$ of the partition.
\footnote
{
For any $\omega\in [0,1)$, this frequency is equal to
$\lim\limits_{l\to\infty}(1/l)\sum_{i=1}^l\chi_1(T^i\omega)=\lambda(\pi_1)$,
where $\chi_1(r)=1$ if $r\in\pi_1$, and $\chi_1(r)=0$, otherwise.
}
Thus for any $\delta>0$ for all sufficiently large $n$, the measure $P$
of all sequences $x$ of length $n$ with portion of ones $\le 2r$
is $\ge 1-\delta$. Since $2rn\le\frac{n}{2}$, we obtain a standard upper
bound for Kolmogorov complexity of any such $x$
\begin{equation} \label{cnk-1}
\frac{K(x)}{n}\le\frac{1}{n}\log\sum\limits_{i=0}^{\lceil 2rn\rceil}{n \choose i}+
\frac{2\log n}{n}\le -3r\log r
\end{equation}
for all sufficiently large $n$. By this inequality and by the approximate
equality (\ref{kol-comp-1}), which holds almost surely,
we obtain an upper bound $H\le -3r\log r\le\epsilon$ for entropy $H$ of the
measure $P$ for all sufficiently small $r$.

Let us prove that an infinite sequence $\alpha$ exists such that
the conclusion of Theorem~\ref{theorem-1} holds. We will define $\alpha$
by induction on steps~$s$ as the union of an increasing sequence of
initial fragments
\begin{equation} \label{alpha-1}
\alpha(0)\subset\dots\subset\alpha(k)\subset\dots
\end{equation}
For all sufficiently large $k$, the Kolmogorov complexity
of the initial fragment $\alpha(k)$ will be small if $k$ is odd,
and complexity of $\alpha(k)$ will be large, otherwise.

Define $\alpha(0)$ to be equal to $\Pi_0$-name of some trajectory of
length $\ge h_0$ such that $d_P(\alpha(0))\le 2$.
Such a trajectory exists by Proposition~\ref{deff-1} (Section~\ref{increase}).
Define $s(-1)=s(0)=0$.

{\it Induction hypotheses.} Let a sequence
$\alpha(0)\subset\dots\subset\alpha(k-1)$ is already defined for some $k>0$.

Suppose that the following properties hold:
\begin{itemize}
\item{(i)}
at some step $s(k-1)$ of the construction the word $\alpha(k-1)$
is equal to an $\Pi_{s(k-1)}$~-~name of a trajectory starting at
some point of the support of this gadget;
\item{(ii)}
this name has length $l(\alpha(k-1))>h_{s(k-1)}$;
\item{(iii)}
$d_P(\alpha(k-1))\le\sigma(h_{s(k-2)})-4$ if $k$ is odd, and
$d_P(\alpha(k-1))\le\sigma(h_{s(k-2)})$ and
$P^{s(k-1)}(\alpha(k-1))>(1/8)P(\alpha(k-1))$ if $k$ is even.
\end{itemize}

Let us consider any odd $k$. Define $a=\alpha(k-1)$.
Pick out a set of all intervals of the
gadget $\Pi_{s-1}$ such that for any trajectory starting in this interval
the following properties hold:
\begin{itemize}
\item{}
$\Pi_{s-1}$-name of this trajectory extends $a$;
\item{}
the frequency of visiting the element $\pi_1$ of the partition is $\le 2r$.
\end{itemize}
For any such $\Pi_{s-1}$-name $\gamma$,
\begin{equation} \label{small-1}
K(\gamma)/l(\gamma)\le -3r\log r\le\epsilon,
\end{equation}
where $r$ is sufficiently small. The proof of (\ref{small-1})
is similar to the proof of (\ref{cnk-1}).
By the ergodic theorem, for all sufficiently large $s$,
total length of all intervals from this set is $\ge (1/2)P(a)$.

Divide all intervals stacking a column of the gadget $\Pi_s$
into two equal parts: upper half and lower half.
Consider only intervals from the lower part. Any trajectory starting
from a point of an interval from the lower part has length $\ge h_s$.
Fix some sufficiently large $s$ as above and define $s(k)=s$.

Let $U_s(a)$ be a set of all intervals from the lower part of the gadget
$\Pi_{s}$ generating trajectories with
$\Pi_{s}$-names extending $a$ and satisfying the inequality~(\ref{small-1}).
By definition the total length of all intervals from $U_{s}(a)$ is more
than $(1/4)P(a)$.
Let $D_a$ be a set of all $\Pi_{s}$-names of these trajectories.

We can also say that the inequality $P^s(\tilde D_a)=P^s(a)>(1/4)P(a)$ holds,
where we denote $\tilde D=\cup_{x\in D}\Gamma_x$ for any finite set of
strings $D$.

It is easy to prove that a set $C_a\subseteq D_a$ exists such that
$P(\tilde C_a)>(1/8)P(\tilde D_a)$ and $P^s(b)>(1/8)P(b)$
for each $b\in C_a$.

By Proposition~\ref{deff-1} an $b\in C_a$ exists such that
$d_P(b^j)\le d_P(a)+4$ for $l(a)\le j\le l(b)$. Define $\alpha(k)=b$.

By induction hypotheses (ii) and (iii) the inequalities
$d_P(a)\le\sigma(h_{s(k-2)})-4$ and $l(a)\ge h_{s(k-1)}>h_{s(k-2)}$ hold.
Then $d_P(b^j)\le\sigma(h_{s(k-2)})\le\sigma(l(a))\le\sigma(j)$ for all
$l(a)\le j\le l(b)$.

Notice, that $l(b)\ge h_{s(k)}$, since any trajectory with a name
$b$ starts from an interval of the lower half of the gadget $\Pi_s$,
and the height of this gadget is $\ge 2h_s$.
Hence, the induction hypotheses (i)-(iii) are valid for the next
step of induction.

The condition (\ref{liminf-1}) is true, since condition~(\ref{small-1})
holds for infinite number of initial fragments $\alpha(k)$
of the sequence $\alpha$.

Let $k$ be even. Put $b=\alpha(k-1)$.
Let $s=s(k-1)+1$. Define $s(k)=s$.

Consider an arbitrary column of the gadget $\Delta_{s-1}$.
Divide all its intervals into two equal parts: upper half and lower half.
Any trajectory starting from an interval of the lower part has
length $\ge M/2$, where $M$ is the height of the gadget $\Delta_{s-1}$.
By the construction $M\ge 2h_{s-1}$.

Total length of all such intervals is equal to
$\frac{1}{2}\lambda(\hat\Delta_{s-1})$.

Let us consider the names $x^{M/2}$ of initial fragments of length
$M/2$ of all these trajectories. By incompressibility property
of Kolmogorov complexity (\ref{incom-1}) and by choice of
$M$ the uniform Bernoulli measure $L$ of all sequences of length
$M/2$ satisfying
$$
\frac{K(x^{M/2})}{l(x^{M/2})}<1-\frac{2}{h_{s-2}},
$$
is less than $2^{-M/h_{s-2}}\le 1/4$. The names of initial fragments
(of length $M/2$) of the rest part of trajectories starting from
intervals of the lower part of the gadget $\Delta_{s-1}$ satisfy
\begin{equation} \label{K-2}
\frac{K(x^{M/2})}{l(x^{M/2})}\ge 1-\frac{2}{h_{s-2}}.
\end{equation}

By definition for any step $s$ of the construction, the equality
$$
P^{s-1}(x)=2^{-l(x)}\lambda(\hat\Delta_{s-1})
$$
holds for the name $x$ of any trajectory of the gadget $\Delta_{s-1}$.
We conclude from this equality that the uniform measure of all intervals
stacked in the lower half of the gadget $\Delta_{s-1}$ and generating
trajectories with names (more correctly, with initial fragments $x^{M/2}$
of such names) satisfying (\ref{K-2}) is at least
$\frac{1}{4}\lambda(\hat\Delta_{s-1})$.

By~(\ref{speed-1})
\begin{eqnarray}\label{vol-2}
\gamma=\frac{\lambda(\hat\Delta'')}{\lambda(\hat\Pi_{s-1})}=
\frac{\lambda(\hat\Delta_{s-1})}{2\lambda(\hat\Pi_{s-1})}
=\frac{2^{-s+1}r}{1-2^{-s+2}}>
\nonumber
\\
>2^{-s+1}r\ge 2^{-(\sigma(h_{s-1})-\sigma(h_{s-2})+12)}.
\end{eqnarray}
Let us consider $R_{s}$-fold independent cutting and stacking
of the gadget $\Pi_{s-1}\cup\Delta''$ in more details. At first,
we cut of this gadget on $R_{s}$ copies. When we stack the next copy on
already defined part of the gadget the portion of all trajectories
of any column from the previously constructed part, which go to a
subcolumn from the gadget $\Delta''$, is equal to
\begin{equation} \label{ratio-1}
\frac{\lambda(\hat\Delta'')}{\lambda(\hat\Pi_{s-1})+\lambda(\hat\Delta'')}
=\frac{\gamma}{1+\gamma}.
\end{equation}
This is true, since by definition any column is covered by a set of
subcolumns with the same distribution
as the gadget $\Pi_{s-1}\cup\Delta''$ has. 

Total length of all intervals of the gadget $\Pi_{s-1}$ such that trajectories
with names extending $b$ start from these intervals is equal to
$P^{s-1}(b)$.

Consider the lower half of all subintervals obtained by cutting and
stacking of the gadget $\Pi_{s-1}$ in which trajectories with
$\Pi_{s-1}$-names extending $b$ start. The length of any such
trajectory is at least $h_s$.
%By this reason some inductive hypothesis will be true.
The measure of all remaining subintervals decreases twice.
After that, we consider
a subset of these subintervals, such that trajectories starting
from subintervals of this subset go into subcolumns of the gadget
$\Delta''$. The measure of the set of remaining subintervals is multiplied
by a factor $\gamma/(1+\gamma)$. Further, consider the subintervals
from the remaining part generating trajectories whose names have in
$\Delta''$ fragments satisfying~(\ref{K-2}). The measure of
the remaining part can be at least $1/4$ of the previously considered
part. This follows from the previous estimate of the portion
of subintervals generating trajectories in the gadget $\Delta''$
of length $\ge M/2$ satisfying~(\ref{K-2}).
%\footnote
%{
%Remember, that $M$ ($\ge 2h_{s-1}$) is the height of gadgets
%$\Pi_{s-1}$, $\Delta_{s-1}$.
%}
Let $D_b$ be a set of all $\Pi_s$-names of all trajectories starting from
subintervals remaining after these selection operations. Then
\begin{equation}\label{inequ-1}
P^s(D_b)\ge\frac{\gamma}{8(1+\gamma)}P^{s-1}(b).
\end{equation}
The name of any such trajectory has an initial fragment of type
$bx'x^{M/2}$, where $x'x^{M/2}$ is the name of a fragment
of this trajectory corresponding to its path in the gadget $\Delta_{s-1}$.
The word $x^{M/2}$ has length $M/2$ and satisfies (\ref{K-2}).
The word $x'$ is the name of a fragment of the trajectory which goes
from an interval of the lower part to an interval generating trajectory
with the name $x^{M/2}$.
We have $l(bx'x^{M/2})\le 2M=4l(x^{M/2})$. By (\ref{K-1}) and (\ref{K-2})
we obtain for these initial fragments of sufficiently large length:
\begin{eqnarray}
\frac{K(bx'x^{M/2})}{l(bx'x^{M/2})}\ge
\frac{K(x^{M/2})-2\log l(bx')}{4l(x^{M/2})}\ge
\frac{1}{4}-\frac{1}{h_{s-2}}.
\label{K-3}
\end{eqnarray}
We have $P^{s-1}(b)>(1/8)P(b)$ by induction hypothesis (iii).
After that, taking into account that $\gamma\le 1$, we deduce from
(\ref{inequ-1})
\begin{eqnarray*}
P(\tilde D_b)\ge P^{s-1}(D_b)\ge\frac{\gamma}{128}P(b).
\end{eqnarray*}
By Proposition~\ref{deff-1} an $c\in D_b$ exists such that
\begin{eqnarray*}
d_{P}(c^j)\le d_{P}(b)+1-\log\frac{\gamma}{128}\le
\nonumber
\\
\le d_{P}(b)+(\sigma (h_{s-1})-\sigma (h_{s-2})-12)+8\le
\nonumber
\\
\le\sigma (h_{s-1})-4=\sigma (h_{s(k-1)})-4
\end{eqnarray*}
for all $l(b)\le j\le l(c)$. Here we have
$d_P(b)\le\sigma (h_{s(k-2)})\le\sigma(h_{s-2})$ by induction hypothesis (iii).
We also used inequality (\ref{vol-2}). Besides, by induction hypothesis (ii)
we have $l(b)\ge h_{s-1}$. Therefore,
$$
d_{P}(c^j)<\sigma(h_{s-1})\le\sigma(l(b))\le\sigma(j)
$$
for $l(b)\le j\le l(c)$. Define $\alpha(k)=c$.

It is easy to see that all induction hypotheses (i)-(iii)
are valid for $\alpha(k)$.

An infinite sequence $\alpha$ is defined by a sequence of initial
fragments (\ref{alpha-1}). We have proved that
$d_{P}(\alpha^j)\le\sigma(j)$ for all $j\ge l(\alpha(1))$.

By the construction there are infinitely many initial fragments
of the sequence $\alpha$ satisfying~(\ref{K-3}). The sequence $h_s$,
where $s=0,1,\dots$, is monotone increased. So, the
condition~(\ref{limsup-1}) hold.
$\bigtriangleup$

We complete this section by the proof of Theorem~\ref{theorem-2}.

Since any universal coding scheme is asymptotically optimal,
i.e., for almost all sequences,
it has compressing ratio (\ref{asym-zv}) equals to Kolmogorov complexity
compressing ratio (\ref{kol-comp-1}), we obtain an instability property for
any universal coding scheme.

In more details, for any $n$ a decoding algorithm $\psi_n$ of the code
$\{\phi_n\}$ is defined by a program of length $\log n+O(1)$. Then we have
an upper bound for Kolmogorov complexity of any initial fragment
of the sequence $\alpha$ defined in the proof of Theorem~\ref{theorem-2}
$$
%\begin{equation}\label{K-phi-1}
K(\alpha^n)\le l(\phi_n(\alpha^n))+O(\log n).
$$
%\end{equation}
Inequality (\ref{limsup-1mb}) follows from the inequality
(\ref{limsup-1}) of Theorem~\ref{theorem-1}. The proof of
inequality~(\ref{liminf-1mb}) is similar to the proof of
inequality~(\ref{liminf-1}) of Theorem~\ref{theorem-1}.
In that proof we replace condition~(\ref{small-1})
on $l(\phi_n(\omega^n))/n\le\epsilon$. We also take into account
the property~(\ref{asym-zv}) of asymptotic optimality
of the code $\{\phi_n\}$.

\section{Proof of Theorem~\ref{th-1}}~\label{sec-2}

To prove Theorem~\ref{th-1} we interpret $T$ as a transformation
of infinite binary sequences preserving the measure $L$ on $\Omega$.
To do this we will use a natural correspondence between points of $[0,1)$
and infinite binary sequences representing dyadic
representations of these points (see Section~\ref{sec-1}).
This correspondence is one-to-one besides a set of measure zero.

We consider also a natural correspondence between finite binary sequences and
subintervals of $[0,1)$ with dyadically rational endpoints.
A finite sequence $\alpha$ of length $n$ corresponds to an interval
$[a,a+2^{-n})$ with dyadically rational endpoints, where
$a=\sum_{1\le i\le n}\alpha_i2^{-i}$. Also, any sequence
of strictly nested intervals with dyadically rational endpoints
corresponds to an infinite binary sequence.

When stacking a subinterval $J$ of $[0,1)$
on a subinterval $I$ we simply add to each
point $\alpha\in I$ a rational number $q$, i.e.,
to define $T\alpha=\alpha+g\in J$.
We can interpreted this transformation $T$ as a computable
operation transforming any binary representation of a number $\alpha$ to
a binary representation of the number $\alpha+q$.
This operation transforms any binary representation $\alpha_1\dots\alpha_n$
of the number $\alpha$ from below up to $2^{-n}$ to a binary representation
of $\alpha+q$ from below up to $2^{-(n-1)}$. We use a sufficiently accurate
binary representation of the rational number $q$.

In what follows we mean by $T$ the corresponding computable operation
on $\Omega$.

We use the construction of Section~\ref{sec-1} to show that
an infinite sequence $\omega\in\Omega$ exists such that
$Km(\omega^n)\ge n-\sigma(n)$ for all $n$ and the limit
(\ref{e-1a}) does not exist for the name
$\chi(\omega)\chi(T\omega)\chi(T^2\omega)\dots$ of its trajectory,
where $\chi(\omega)=i$ if $\omega\in\pi_i$, $i=0,1$. 

We will prove that an $\omega\in\Omega$ exists such that
$Km(\omega^n)\ge n-\sigma(n)$ for all $n$ and the limit
(\ref{e-1a}) does not exist for the name
$\chi(\omega)\chi(T\omega)\chi(T^2\omega)\dots$ of its trajectory,
where $\chi(\omega)=i$ if $\omega\in\pi_i$, $i=0,1$. More precise,
we prove that
\begin{eqnarray}
\limsup\limits_{n\to\infty}\frac{1}{n}\sum\limits_{i=0}^{n-1}\chi(T^i\omega)\ge
1/16,
\label{iineq-1}
\\
\liminf\limits_{n\to\infty}\frac{1}{n}
\sum\limits_{i=0}^{n-1}\chi(T^i\omega)\le 2r,
\label{iineq-2}
\end{eqnarray}
where $r$ is sufficiently small and an indicator function $\chi$
is defined below.
\footnote{We suppose
that the rational number $r$ from the definition of the partition
$\pi_0$, $\pi_1$ has a dyadic rational form.}

We will define by induction on steps~$s$ a sequence $\omega$ satisfying
the conclusion of Theorem~\ref{th-1} as the union of an
increasing sequence of initial fragments
\begin{equation} \label{alpha-1s}
\omega(0)\subset\dots\subset\omega(k)\subset\dots
\end{equation}
Using Proposition~\ref{deff-1} (Section~\ref{increase}), where $P$
is the uniform measure $\lambda$,
define $\omega(0)$ such that $d(\omega(0)^j)\le 2$
for all $j\le l(\omega(0))$. Define $s(-1)=s(0)=0$.

{\it Induction hypotheses.} Suppose that $k>0$ and a sequence
$\omega(0)\subset\dots\subset\omega(k-1)$ is already defined. Also,
the interval with dyadically rational endpoints corresponding to
$\omega(k-1)$ is a subset of an interval
from the support of the gadget $\Pi_{s(k-1)}$. We suppose that
\begin{itemize}
\item{(i)}
$l(\omega(k-1))>h_{s(k-1)}$;
\item{(ii)}
$d(\omega(k-1))\le\sigma(h_{s(k-2)})-3$
if $k$ is odd and $d(\omega(k-1))\le\sigma(h_{s(k-2)})$ if $k$ is even,
where $d(\omega^n)=n-Km(\omega^n)$ is the deficiency of randomness.
\end{itemize}

Consider an odd $k$. Denote $a=\omega(k-1)$.

Let us consider a sufficiently large $s$ and a set of all finite strings
extending $a$ such that the corresponding intervals with dyadically rational
endpoints are subsets of intervals from columns of the gadget $\Pi_{s-1}$ and
generate $\Pi_{s-1}$-trajectories with frequency of visiting the element
$\pi_0$ of the partition less than $2r$.
By the ergodic theorem the total measure of these subintervals tends
to $2^{-l(a)}$ as $s\to\infty$.

We consider a sufficiently large $s$ such that
the total measure of the set of all intervals
with dyadically rational endpoints locating in the lower half
of the gadget $\Pi_s$ is $\ge (1/4)2^{-l(a)}$. Let $C_a$ be a
set of strings corresponding to these subintervals.

Fix some such $s$ and define $s(k)=s$.

By Proposition~\ref{deff-1} (Section~\ref{increase}), where $P=L$,
an $b\in C_a$ exists such that
$d(b^j)\le d(a)+3$ for each $l(a)\le j\le l(b)$. Define $\omega(k)=b$.
By induction hypotheses (i)-(ii) inequalities
$d(a)\le\sigma(h_{s(k-2)})-3$ and $l(a)\ge h_{s(k-1)}>h_{s(k-2)}$ hold.
Then $d(b^j)\le\sigma(h_{s(k-2)})\le\sigma(l(a))\le\sigma(j)$ for all
$l(a)\le j\le l(b)$. Also, we can take $l(b)\ge h_{s(k)}$.
Therefore, the induction hypotheses and condition (\ref{iineq-2}) are valid
for the next step of induction.

Let $k$ be even. Put $b=\omega(k-1)$ and $s=s(k-1)+1$. Define $s(k)=s$.

Let us consider an arbitrary column of the gadget $\Delta_{s-1}$.
Divide all its intervals into two equal parts: upper half and lower half.
Any interval of the lower part generates a trajectory of
length $\ge M/2$, where
$M\ge 2h_{s-1}$ is the height of the gadget $\Delta_{s-1}$.
The uniform measure of all such intervals is equal to
$\frac{1}{2}\lambda(\hat\Delta_{s-1})$.

By Hoeffding inequality (\ref{strong-1g}) the measure $\lambda$
of all points of support of the gadget $\Delta_{s-1}$ whose
trajectories have length $\ge M/2$ and frequency of ones $\le 1/4$
is less than
$$
2^{-\frac{1}{16}M}\lambda(\hat\Delta_{s-1})\le
\frac{1}{4}\lambda(\hat\Delta_{s-1})
$$
(assume that $M$ is sufficiently large).
Then all such points from intervals of the lower part of the gadget
$\Delta_{s-1}$ whose trajectories $\alpha$ have length $\ge M/2$
and frequency of ones more than $3/4$
have total measure at least $\frac{3}{4}\lambda(\hat\Delta_{s-1})$.

By~(\ref{speed-1})
\begin{eqnarray}\label{vol-2s}
\gamma=\frac{\lambda(\hat\Delta'')}{\lambda(\hat\Pi_{s-1})}=
\frac{\lambda(\hat\Delta_{s-1})}{2\lambda(\hat\Pi_{s-1})}
=\frac{2^{-s+1}r}{1-2^{-s+2}}>
\nonumber
\\
>2^{-s+1}r\ge 2^{-(\sigma(h_{s-1})-\sigma(h_{s-2})+12)}.
\end{eqnarray}

Let $\Phi_s$ is the gadget generated by cutting and stacking of
$\Pi_{s-1}\cup\Delta''$.

Consider a set of all binary strings correspondent to intervals from
the lower part of $\Phi_s$ such that trajectories starting from these
intervals and extending the sequence $b$ pass through an upper subcolumn of the 
gadget $\Delta''$ and have frequencies of ones at least $1/4$. 
Notice that any copy of the gadget $\Delta''$ has the same frequency
characteristics of trajectories.

%For any such subinterval
%consider a maximal subinterval with dyadically rational endpoints
%containing in it. 

Let $D_b$ be a set of all binary strings correspondent to these intervals.
By definition trajectory of any such interval has length at most $2M$ and some 
initial fragment of its name has at least $M/8$ ones. Hence, frequency of ones
in the name of any such trajectory is at least $\frac{1}{16}$.

Since $\gamma\le 1$, the total measure of all such intervals is
at least $\frac{\gamma}{32}2^{-l(b)}$. By
Proposition~\ref{deff-1}, where $P=\lambda$, an $c\in D_b$ exists such that
\begin{eqnarray*}
d(c^j)\le d(b)+1-\log\frac{\gamma}{32}\le
\\
\le
d(b)+(\sigma (h_{s-1})-\sigma (h_{s-2})-12)+6\le
\\
\sigma (h_{s-1})-6<\sigma (h_{s(k-1)})-3
\end{eqnarray*}
for all $j$ such that $l(b)\le j\le l(c)$. Here we have
$d(b)\le\sigma (h_{s(k-2)})\le\sigma(h_{s-2})$ by induction hypothesis (ii).
We also have used the inequality (\ref{vol-2s}). Besides, by the induction
hypothesis (i) we have $l(b)\ge h_{s-1}$. Therefore,
$$
d(c^j)<\sigma(h_{s-1})\le\sigma(l(b))\le\sigma(j)
$$
for all $j$ such that $l(b)\le j\le l(c)$. Define $\omega(k)=c$.
It is easy to see that induction hypotheses (i)-(ii) are valid for $\omega(k)$.

An infinite sequence $\omega$ is defined by a sequence of its initial
fragments (\ref{alpha-1s}). We have proved that
$d(\omega^j)\le\sigma(j)$ for all $j$.

By the construction there are infinitely many initial fragments
of trajectory of the sequence $\omega$ with frequency of ones $\ge 1/16$
in their names. Hence, the condition~(\ref{iineq-1}) holds.
$\bigtriangleup$

The proof of Theorem~\ref{th-1b} is more complicated. Consider a sequence 
of pairwise disjoint subsets $\Gamma_{x_i}$,
where $x_i=1^{i-1}0$, such that $L(\cup_i\Gamma_{x_i})=1$.
For any $\Gamma_{x_i}$, we apply the construction of Section~\ref{sec-1} 
to a computable sequence $\sigma_i(n)$ of all
partial recursive functions and define a computable ergodic measure preserving
transformation $T_i$ on $\Gamma_{x_i}$ for each $i$. 
The needed transformation is defined as union of all these transformations 
$T_i$. We pass details of this construction.

\appendix

\section{Auxiliary notions and assertions}\label{app}

\subsection{Algorithmic randomness}\label{random-1}

Let $\Theta=\{0,1\}^*$ be a set of all finite binary sequences
(strings) and $\Omega=\{0,1\}^\infty$ be a set of all infinite
binary sequences. Let $l(\alpha)$ denotes the length of a
sequence $\alpha$ ($l(\alpha)=\infty$ for $\alpha\in\Omega$).

For any finite or infinite sequence $\omega=\omega_1\omega_2\dots$, we write
$\omega^n=\omega_1\omega_2\dots\omega_n$, where $n\le l(\omega)$.
Also, we write $\alpha\subseteq\beta$ if $\alpha=\beta^n$ for some $n$.
Two finite sequences $\alpha$ and $\beta$ are incomparable if
$\alpha\not\subseteq\beta$ and $\beta\not\subseteq\alpha$.
A set $A\subseteq\Theta$ is prefix-free if any two sequences from $A$
are incomparable.

A complexity of a word $x\in\Theta^*$
(with respect to a word $y\in\Theta^*$) is equal to the length of the shortest
binary codeword $p$ (i.e. $p\in\{0,1\}^*$) by which given $y$
the word $x$ can be reconstructed:
$$
K_\psi(x|y)=\min\{l(p):\psi(p,y)=x\}.
$$
We suppose that $\min\emptyset=+\infty$.

By this definition the complexity of $x$ depends on a computable
(partial recursive) function
$\psi$ -- method of decoding. Kolmogorov proved that {\it an
optimal decoding algorithm} $\psi$ exists such that 
$$
K_\psi(x|y)\le K_{\psi'}(x|y)+O(1)
$$
holds for any computable decoding function $\psi'$ and for all words
$x$ and $y$. 
%Here $K(\psi')$ is the length of the shortest program computing values of $\psi'$.
%\footnote
%{
%We suppose that some universal programming language is fixed,
%and all decoding programs are written in this language
%(the constant $O(1)$ depends on this language).
%}

We fix some optimal decoding function $\psi$. The value
$K(x|y)=K_\psi(x|y)$ is called (conditional) Kolmogorov complexity
of $x$ given $y$. Unconditional complexity of $x$ is defined
$K(x)=K(x|\Lambda)$.

%It can be proved that a corresponding to $\psi$ coding algorithm
%computing by $x$ a
%codeword $p$ of minimal length such that $\psi(p)=x$ does not exist.

We will use some properties of Kolmogorov complexity from
Li and Vitanyi~\cite{LiV97}.
Incompressibility property asserts that for any positive integer numbers
$n$ and $m$ a portion of all sequences $x$ of length $n$ such that
\begin{equation}\label{incom-1}
K(x)<n-m,
\end{equation}
is less than $2^{-m}$. Indeed, the number of all $x$ satisfying
this inequality does not exceed the number of all binary programs
generating them. Since the length of any such program is less than $n-m$
the number of these programs is less than $2^{n-m}$.

Let $x$ and $b$ be finite words. It is easy to construct
a function which given any program computing $bx$ and the length of $b$
computes the word $x$. Therefore, we obtain an extension property
\footnote
{
Recall that we consider in the following logarithms on the base 2.
}
\begin{equation} \label{K-1}
K(x)\le K(bx)+2\log l(b)+c
\end{equation}
for any $x$, where $c$ is a positive constant not depending from
$b$ and $x$.

We use a probability space $(\Omega,{\cal P},P)$,
where $\Omega=\{0,1\}^\infty$ is the set of all infinite binary sequences
and $\cal P$ is the set of all Borel subsets $\Omega$.
A probability measure $P$ on $\Omega$ is defined
as follows: define $P(\Gamma_\alpha)$ for each
finite sequence $\alpha$, where
$\Gamma_\alpha=\{\omega\in\Omega:\alpha\subset\omega\}$, and extend
it for all Borel subsets of $\Omega$.

Let $\cal R$ be a set of all real numbers, $\cal Q$ be a set of
all rational numbers.

In this paper we consider computable measures on $\Omega$.
A function $f:\Theta\to {\cal R}$ is called computable if there
exists an algorithm which given $x\in\Theta$ and a rational $\epsilon>0$
computes a rational approximation of a number $f(x)$ with
accuracy $\epsilon$.

An important example of computable measure is the uniform Bernoulli measure.
The uniform Bernoulli measure on $\Omega$ is defined
$$
L(\Gamma_\alpha)=2^{-l(\alpha)}
$$ 
for any finite binary sequence $\alpha$.

An open subset $U$ of $\Omega$ is called effectively open if it
can be represented as a union of a computable sequence of
intervals: $U=\cup_{i=1}^\infty\Gamma_{\alpha_i}$, where
$\alpha_i=f(i)$ is a computable function from $i$. A sequence
$U_n$, $n=1,2,\dots$ of effectively open sets is called
effectively enumerable if it can be represented as
$U_n=\cup_{i=1}^\infty\Gamma_{\alpha_{n,i}}$, where
$\alpha_{n,i}=f(n,i)$ is a computable function from $n$ and
$i$.

Martin-L\"of test of randomness with respect to a computable measure $P$
is an effectively enumerable sequence $U_n$, $n=1,2,\dots$, of effectively
open sets such that $P(U_n)\le 2^{-n}$ for all $n$. If the real number
$P(U_n)$ is computable then the test $U_n$, $n=1,2,\dots$, is called
Schnorr test of randomness.

An infinite binary sequence $\omega$ is called  Martin-L\"of random
with respect to $P$ if $\omega\not\in\cap U_n$ for any Martin-L\"of
test of randomness $U_n$, $n=1,2,\dots$. A notion of Schnorr random
sequence is defined analogously.

An equivalent definition of randomness can be obtained using
Solovay tests of randomness.
A computable sequence $\{x_n:n=1,2,\dots\}$ of finite strings is called
Solovay test of randomness if the series
$\sum\limits_{n=1}^\infty 2^{-l(x_n)}$ converges.

An infinite sequence $\omega$ passes a Solovay test of randomness 
$\{x_n:n=1,2,\dots\}$ if $x_n\not\subseteq\omega$ for almost all $n$.

We have used an equivalence between 
Martin-L\"of and Solovay tests of randomness.
\begin{proposition}\label{Sol-Mart-1}
An infinite sequence $\omega=\omega_1\omega_2\dots$ is Martin-L\"of random
if and only if it passes each Solovay test of randomness.
\end{proposition}
{\it Proof}. Assume that $\omega$ is not Martin-L\"of random.
Then a Martin-L\"of test $U_n$, $n=1,2,\dots$, exists such that
$\omega\in\cap U_n$.
Define a Solovay test of randomness as follows. Since $L(U_n)\le 2^{-n}$ 
for all $n$, we can effectively compute a prefix-free sequence 
of finite strings $x_n$, $n=1,2,\dots$, such that 
$\cup_n\Gamma_{x_n}=\cup_n U_n$ and the series
$\sum\limits_{n=1}^\infty 2^{-l(x_n)}$ converges. 
Then $x_n\subset\omega$ for infinitely many $n$.

On the other side, assume that for some Solovay test $x_n$, $n=1,2,\dots$,
$x_n\subset\omega$ for infinitely many $n$. Let 
$\sum\limits_{n=1}^\infty 2^{-l(x_n)}<2^K$, where $K$ is a positive 
integer number. Let $U_n$ be a set of all infinite $\omega$ such that
$|\{m:x_m\subset\omega\}|\ge 2^{n+K}$. It is easy to verify that $U_n$
is a Martin-L\"of test of randomness and that $\omega\in\cap U_n$.
$\triangle$

We strengthens the definition of Solovay test of randomness -- we
consider total Solovay tests of randomness.

A series $\sum\limits_{i=1}^\infty r_i$ converges with a computable rate
of convergence if a computable function $m(\delta)$ exists such that
%\begin{eqnarray}
$|\sum\limits_{i=m(\delta)}^\infty r_i|\le\delta$
%\label{coverg-1}
%\end{eqnarray}
for each positive rational number $\delta$.

A computable sequence $\{x_n:n=1,2,\dots\}$ of finite strings is called
total Solovay test of randomness if the series
$\sum\limits_{n=1}^\infty 2^{-l(x_n)}$ converges with
computable rate of convergence.

\begin{proposition}\label{Sol-Mart-11}
An infinite sequence $\omega=\omega_1\omega_2\dots$ is Schnorr random
if and only if it passes each total Solovay test of randomness.
\end{proposition}
The proof is similar to the proof of Proposition~\ref{Sol-Mart-1}.

Using some modification of decoding algorithms, we obtain an equivalent
definition of the notion of algorithmic random sequence in terms of
algorithmic complexity (see Li and Vitanyi~\cite{LiV97}).

Let us define a notion of a monotonic computable transformation of
binary sequences. A computable representation of such an operation is
a set $\hat\psi\subseteq \{0,1\}^*\times \{0,1\}^*$ such that
\begin{itemize}
\item{(i)}
the set $\hat\psi$ is recursively enumerable;
\item{(ii)}
for any $(x,y), (x',y')\in\hat\psi$ if $x\subseteq x'$ then
$y\subseteq y'$ or $y'\subseteq y$;
\item{(iii)}
if $(x,y)\in\hat\psi$ then $(x,y')\in\hat\psi$ for all $y'\subseteq y$.
\end{itemize}
The set $\hat\psi$ defines a monotonic with respect to $\subseteq$
decoding function
\footnote
{
Here the by supremum we mean an union of all comparable $x$ in one sequence.
}
\begin{equation}\label{mon-cod}
\psi(p)=\sup\{x:\exists p'(p'\subseteq p\&(p',x)\in\hat{\psi})\}.
\end{equation}
Any computable monotonic function $\psi$ determines the
corresponding measure of complexity
$$
Km_\psi(x)=\min\{l(p):x\subseteq\psi(p)\}=\min\{l(p):(x,p)\in\hat\psi\}.
$$
The invariance property also holds for monotonic measures of
complexity: an optimal computable operation $\psi$ exists such
that $Km_\psi(x)\le Km_{\psi'}(x)+O(1)$ for all computable
operations $\psi'$ and for all finite binary sequences $x$.

The corresponding optimal complexity $Km(x)=Km_\psi(x)$ differs from
simple Kolmogorov complexity $K(x)$ by a term of order $O(\log l(x))$
$$
K(x)\le Km(x)+O(1)\le K(x)+O(\log l(x))
$$
An infinite sequence $\omega$ is Martin-L\"of random with
respect to a computable measure $P$ if and only if
\begin{eqnarray}\label{M-r-1}
Km(\omega^n)=-\log P(\omega^n)+O(1)
\end{eqnarray}
as $n\to\infty$.
In particular, an infinite binary sequence
$\omega$ is Martin-L\"of random
(with respect to uniform measure) if and only if
$Km(\omega^n)=n+O(1)$ as $n\to\infty$
(see for details Li and Vitanyi~\cite{LiV97}).

A function
$$
d_P(\omega^n)=-\log P(\omega^n)-Km(\omega^n)
$$
is called universal
deficiency of randomness (with respect to a measure $P$). For the 
uniform measure, $d(\omega^n)=n-Km(\omega^n)$.

By (\ref{M-r-1}), $d_P(\omega^n)<\infty$ if and only if a sequence $\omega$ is
Martion-L\"of random with respect to $P$.

A transformation $T$ of the set $\Omega$ is computable if a computable
representation $\hat\psi$ exists such that (i)-(iii) hold and
\begin{equation}\label{mon-cod-1}
T(\omega)=\sup\{y:x\subseteq\omega\&(x,y)\in\hat{\psi})\}
\end{equation}
for all infinite $\omega\in\Omega$.

Recall that $\cal Q$ denotes the set of all rational numbers, and let $r$
denotes an arbitrary element of $\cal Q$.

%A function $f:\Theta\to {\cal R}$ is called computable if there
%exists an algorithm which given a rational number $\epsilon>0$
%and a finite binary sequence $x$ outputs a raional number $r$
%such that $|f(x)-r|<\epsilon$.

A nonempty set is called recursively enumerable if it is a range of some
computable function defined on the set of positive integer numbers.

Let $f$ be a function from $\Omega$ to ${\cal R}\cup\{-\infty,+\infty\}$.
A function $\hat f:\Theta\to {\cal Q}$ is called lower approximation of $f$
if $r<f(\omega)$ if and only if $r<\hat f(x)$ for some $x\subset\omega$.

A function $f$ is lower semicomputable is it has a computable lower 
approximation. This definition means that if $r<f(\omega)$ this fact 
will sooner or later be learned by some algorithm.

A function $f$ is called upper semicomputable if the function $-f$
is lower semicomputable.
\footnote{An equivalent definition can be defined in terms of computable 
upper approximation.}  
A function $f:\Omega\to{\cal R}\cup\{-\infty,+\infty\}$ is computable 
if it is lower and upper semicomputable.

\subsection{Bounded increase of deficiency of randomness}\label{increase}

In the proof of Theorem~\ref{theorem-1} a property of
a bounded increase of the deficiency of randomness was used.
Let $P$ be a measure, $P(x)\not = 0$ and a set $A$ consists of words $y$
such that $x\subseteq y$.

Let $\tilde A=\cup\{\Gamma_y:y\in A\}$ for any $A\subseteq\{0,1\}^*$.
Define $P(\tilde A|x)=P(\tilde A)/P(x)$.
\begin{proposition}  \label{deff-1}
Let $P$ be a computable measure, $x$ be a word, $P(x)\not = 0$ and a set $A$ consists
of words $y$ such that $x\subseteq y$ and $P(\tilde A)>0$. Then for any
$0<\mu<1$ a subset $A'\subseteq A$ exists such that
$P(\tilde A')>\mu P(\tilde A)$ and
$$
d_P(y^n)\le d_P(x)-\log(1-\mu)-\log P(\tilde A|x)
$$
for all $y\in A'$ and $l(x)\le n\le l(y)$.
\end{proposition}
{\it Proof}. We will use in the proof a notion of supermartingale
\cite{Shi80}.
A function $M$ is called $P$--supermartingale if it is defined
on $\{0,1\}^*$ and satisfies conditions:
\begin{itemize}
\item{}
$M(\Lambda)\leq 1$;
\item{}
$M(x)\geq M(x0)P(0|x)+M(x1)P(1|x)$ for all $x$, 
where $P(\nu|x)=P(x\nu)/P(x)$ for $\nu=0, 1$ (we put here
$0/0=0*\infty=0$).
\end{itemize}
A supermartingale $M$ is lower semicomputable if the set $\{(r,x):r<M(x)\}$,
where $r$ is a rational number, is a range of some computable function.
We will consider only nonnegative supermartingales.

Let us prove that the deficiency of randomness is bounded by a logarithm
of some lower semicomputable supermartingale.
\begin{lemma} Let $P$ be a computable probability measure. Then there
exists a lower semicomputable $P$--supermartingale $M$ such that
$d_P(x)\le\log M(x)$ for all $x$.
\end{lemma}
{\it Proof}. Let some optimal function $\psi$ satisfying
(\ref{mon-cod}) defines the monotone complexity $Km(x)$. Define
\begin{equation} \label{semi-1}
Q(\alpha)=L(\cup\{\Gamma_p:\alpha\subseteq\psi(p)\}),
\end{equation}
where $L$ is the uniform Bernoulli measure. It is easy to verify that
$Q(\Lambda)\le 1$ and $Q(\alpha)\ge Q(\alpha0)+Q(\alpha1)$
for all words $\alpha$. Then the function $M(\alpha)=Q(\alpha)/P(\alpha)$
is a $P$--supermartingale.

Since for any $\alpha$ the shortest $p$ such that
$\alpha\subseteq\psi(p)$ is an element of the set from~(\ref{semi-1}),
we have inequality $Q(\alpha)\ge 2^{-Km(\alpha)}$, and so,
$d_P(\alpha)\le\log M(\alpha)$.
$\bigtriangleup$

We have $d_P(x)\le\log M(x)$, where $M$ is lower semicomputable
$P$-supermartingal. Define a set
\begin{eqnarray*}
A_1=\{y\in A : \exists j (l(x)\le j\le l(y)\&M(y^j)>M(x)/((1-\mu)P(A|x)))\}.
\end{eqnarray*}
%A set of words $B$ is called prefix free if for any two distinct
%words $x,y\in B$ conditions $x\not\subseteq y$ and $y\not\subseteq x$ hold.

By definition of supermartingale for any prefix free set $B$
such that $x\subseteq y$ for all $y\in B$ inequality
\begin{equation}  \label{super-1}
M(x)\ge\sum\limits_{y\in B}M(y)P(y|x)
\end{equation}
holds. For any $y\in A_1$ let $y^p$ be the initial fragment of $y$
of maximal length such that $\frac{M(y^p)}{M(x)}>\frac{1}{(1-\mu)P(A|x)}$.
The set $\{y^p : y\in A_1\}$ is prefix free. Then by (\ref{super-1}) we have
\begin{eqnarray*}
1\ge\sum\limits_{y\in A_1}\frac{M(y^p)}{M(x)}P(y^p|x)>
\frac{1}{(1-\mu)P(\tilde A|x)}\sum\limits_{y\in A_1}P(y^p|x)\ge
\nonumber
\\
\ge\frac{1}{(1-\mu)P(\tilde A|x)}P(\tilde A_1|x).
\end{eqnarray*}
From this we obtain $P(\tilde A_1|x)<(1-\mu)P(\tilde A|x)$.
Define
$$
A'=A-\{y\in A : z\subseteq y\mbox { for some }\mbox z\in A_1\}.
$$
Then $P(\tilde A'|x)>\mu P(\tilde A|x)$. For any $y\in A'$ we have
$$
M(y^j)\le M(x)\frac{1}{(1-\mu)P(\tilde A|x)}
$$
for all $l(x)\le j\le(y)$. The proposition follows
from the inequality $d_P(x)\le\log M(x)$.
$\bigtriangleup$

\subsection{Method of cutting and stacking}\label{cat-sta}

An arbitrary measurable mapping of the a probability space
into itself is called a transformation.
A transformation $T$ preserves a measure $P$ if
$P(T^{-1}(A))=T(A)$ for all measurable subsets $A$ of the space.
A subset $A$ is called invariant with respect to $T$ if $T^{-1}A=A$.
A transformation $T$ is called ergodic if each invariant with respect
to $T$ subset $A$ has measure~0~or~1.

Recall the main notions and properties of cutting and stacking method
(see Shields~\cite{Shi91, Shi93}).
A column is a sequence $E=(L_1,\dots,L_h)$ of pairwise disjoint
subintervals of $[0,1)$ of equal width (and with rational endpoints);
$L_1$ is the base, $L_h$ is the top of the column,
${\hat E}=\cup_{i=1}^{h}L_i$ is the support of the
column, $w(E)=\lambda(L_1)$ is the width of the column,
$h$ is the height of the column,
$\lambda({\hat E})=\lambda(\cup_{i=1}^{h}L_i)$ is the measure of the column.

Any column defines an algorithmically effective transformation $T$
which linearly transforms $L_j$ to $L_{j+1}$
for all $j=1,\dots, h-1$.
This transformation $T$ is not defined outside all intervals of the column
and at all points of the top $L_h$ interval of this column.
Denote $T^0\omega=\omega$, $T^{i+1}\omega=T(T^i\omega)$.
For any $1\le j<h$ an arbitrary point $\omega\in L_j$ generates a finite
trajectory $\omega, T\omega,\dots, T^{h-j}\omega$.

Since all points of the interval $L_j$ generate the identical trajectories,
we refer to this trajectory as to trajectory generated by the
interval $L_j$.

A partition $\pi=(\pi_1,\dots,\pi_k)$ is compatible with a column $E$
if for each $j$ there exists an $i$ such that $L_j\subseteq\pi_i$.
This number $i$ is called the name of the interval $L_j$, and the
corresponding sequence of names of all intervals of the column is
called the name of the column $E$.
For any point $\omega\in L_j$, where $1\le j<h$, by $E$--name of
the trajectory $\omega, T\omega,\dots, T^{h-j}\omega$ we mean
a sequence of names of intervals $L_j,\dots, L_h$ from the column $E$.
The length of this sequence is $h-j+1$.

A gadget is a finite collection of disjoint columns.
The width of the gadget $w(\Upsilon)$ is the sum of the widths of its
columns. A union of gadgets $\Upsilon_i$ with disjoint supports
is the gadget $\Upsilon=\cup\Upsilon_i$ whose columns are the columns
of all the $\Upsilon_i$. The support of the gadget $\Upsilon$ is
the union $\hat\Upsilon$ of the supports of all its columns.
A transformation $T(\Upsilon)$ is associated with a gadget $\Upsilon$ if
it is the union of transformations defined on all columns of $\Upsilon$.
With any gadget $\Upsilon$ the corresponding set of finite trajectories
generated by points of its columns is associated. By $\Upsilon$-name
of a trajectory we mean its $E$-name, where $E$ is that column
of $\Upsilon$ to which this trajectory corresponds.
A gadget $\Upsilon$ extends a column $\Lambda$ if the support of
$\Upsilon$ extends the support of $\Lambda$, the transformation
$T(\Upsilon)$ extends the transformation $T(\Lambda)$ and the partition
corresponding to $\Upsilon$ extends the partition corresponding to $\Lambda$.

The cutting and stacking operations that are common used will now be
defined.
The distribution of a gadget $\Upsilon$ with columns
$E_1,\dots,E_n$ is a vector of probabilities
\begin{eqnarray}
\left(\frac{w(E_1)}{w(\Upsilon)},\dots,\frac{w(E_n)}{w(\Upsilon)}\right).
\label{gad-division}
\end{eqnarray}
A gadget $\Upsilon$ is a copy of a gadget $\Lambda$ if they have the same
distribution and the corresponding columns have the same partition names.
A gadget $\Upsilon$ can be cut into $M$ copies of itself
$\Upsilon_m, m=1,\dots, M$, according to a given probability vector
$(\gamma_1,\dots,\gamma_M)$ of type (\ref{gad-division}) by cutting each column
$E_i=(L_{i,j}: 1\le j\le h(E_i))$ (and its intervals) into disjoint
subcolumns $E_{i,m}=(L_{i,j,m}: 1\le j\le h(E_i))$ such that
$w(E_{i,m})=w(L_{i,j,m})=\gamma_m w(L_{i,j})$.
The gadget $\Upsilon_m=\{E_{i,m}:1\le i\le L\}$ is called the copy of
the gadget $\Upsilon$ of width $\gamma_m$. The action of the gadget
transformation $T$ is not affected by the copying operation.

Another operation is the stacking gadgets onto gadgets.
At first we consider the stacking of columns onto columns and
the stacking of gadgets onto columns.

Let $E_1=(L_{1,j}:1\le j\le h(E_1))$ and $E_2=(L_{2,j}:1\le j\le h(E_2))$
be two columns of equal width whose supports are disjoint.
The new column $E_1*E_2=(L_j:1\le j\le h(E_1)+h(E_2))$ is defined as
$L_j=L_{1,j}$ for all $1\le j\le h(E_1)$ and $L_j=L_{2,j-h(E_1)+1}$ for all
$h(E_1)\le j\le h(E_1)+h(E_2)$.
Let a gadget $\Upsilon$ and a column $E$ have the same width, and their
supports are disjoint. A new gadget $E*\Upsilon$ is defined as follows.
Cut $E$ into subcolumns $E_i$ according to the distribution of the gadget
$\Upsilon$ such that $w(E_i)=w(U_i)$, where $U_i$ is the $i$-th column
of the gadget $\Upsilon$. Stack $U_i$ on the top of $E_i$ to get the new
column $E_i*U_i$. A new gadget consists of the columns $(E_i*U_i)$.

Let $\Upsilon$ and $\Lambda$ be two gadgets of the same width
and with disjoint supports. A gadget $\Upsilon*\Lambda$ is defined as
follows. Let the columns of $\Upsilon$ are $(E_i)$. Cut $\Lambda$ into
copies $\Lambda_i$ such that $w(\Lambda_i)=w(E_i)$ for all $i$.
After that, for each $i$ stack the gadget $\Lambda_i$ onto
column $E_i$, i.e. we consider a gadget $E_i*\Lambda_i$.
The new gadget is the union of gadgets $E_i*\Lambda_i$ for all $i$.
The number of columns of the gadget $\Upsilon*\Lambda$ is the product
of the number of columns of $\Upsilon$ on the number of columns of
$\Lambda$.

The $M$-fold independent cutting and stacking of a single gadget
$\Upsilon$ is defined by cutting $\Upsilon$ into $M$ copies $\Upsilon_i$,
$i=1,\dots,M$, of equal width and successively independently cutting
and stacking them to obtain $\Upsilon^{*(M)}=\Upsilon_1*\dots*\Upsilon_M$.

A sequence of gadgets $\{\Upsilon_m\}$ is complete if
\begin{itemize}
\item{}
$\lim\limits_{m\to\infty} w(\Upsilon_m)=0$;
\item{}
$\lim\limits_{m\to\infty} \lambda({\hat\Upsilon}_m)=1$;
\item{}
$\Upsilon_{m+1}$ extends $\Upsilon_m$ for all $m$.
\end{itemize}
Any complete sequence of gadgets $\{\Upsilon_s\}$ determines a transformation
$T=T\{\Upsilon_s\}$ which is defined on interval $[0,1)$ almost surely.

By definition $T$ preserves the measure $\lambda$. In \cite{Shi91}
the conditions sufficient a process $T$ to be ergodic
were suggested. Let a gadget $\Upsilon$ is constructed by cutting and
stacking from a gadget $\Lambda$.
Let $E$ be a column from $\Upsilon$ and $D$ be a column from $\Lambda$.
Then ${\hat E}\cap{\hat D}$ is defined as the union of subcolumns from
$D$ of width $w(E)$ which were used for construction of $E$.

Let $0<\epsilon<1$.
A gadget $\Lambda$ is $(1-\epsilon)$-well-distributed in $\Upsilon$ if
\begin{equation}
\sum_{D\in\Lambda}\sum_{E\in\Upsilon}|\lambda({\hat E}\cap{\hat D})-
\lambda({\hat E})\lambda({\hat D})|<\epsilon.
\end{equation}
We will use the following two lemmas.
\begin{lemma} \label{well-def}
(\cite{Shi91}, Corollary 1), (\cite{Shi93}, Theorem A.1).
Let $\{\Upsilon_n\}$ be a complete sequence of gadgets and for each $n$
the gadget $\{\Upsilon_n\}$ is $(1-\epsilon_n)$-well-distributed in
$\{\Upsilon_{n+1}\}$, where $\epsilon_n\to 0$. Then $\{\Upsilon_n\}$
defines the ergodic process.
\end{lemma}
\begin{lemma} \label{M-fold} (\cite{Shi93}, Lemma 2.2).
For any $\epsilon>0$ and any gadget $\Upsilon$ there is an $M$ such that
for each $m\ge M$ the gadget $\Upsilon$ is $(1-\epsilon)$-well-distributed
in the gadget $\Upsilon^{*(m)}$ constructed from $\Upsilon$ by
$\mbox m$-fold independent cutting and stacking.
\end{lemma}

%\section*{Acknowledgement}

%This research was partially supported by Russian foundation for
%fundamental research: 09-07-00180-a.

\end{document}